\pdfoutput=1
\documentclass[pdftex]{pasj00}

\SetRunningHead{Miura et al.}{X-ray Flare from HD\,160184}
\Received{//}
\Accepted{//}

\begin{document}
\title{Suzaku Detection of an Intense X-Ray Flare\\ from an A-type Star HD\,161084}
\author{
Junichiro~\textsc{Miura},\altaffilmark{1} Masahiro~\textsc{Tsujimoto},\altaffilmark{2,3}
Yohko~\textsc{Tsuboi},\altaffilmark{1} Yoshitomo~\textsc{Maeda},\altaffilmark{4}
Yasuharu~\textsc{Sugawara},\altaffilmark{1} Katsuji~\textsc{Koyama},\altaffilmark{5} \&
Shigeo~\textsc{Yamauchi}\altaffilmark{6}
}
\altaffiltext{1}{Department of Physics, Faculty of Science \& Engineering, Chuo
University, 1-13-27 Kasuga, Bunkyo, Tokyo 112-8551}
\altaffiltext{2}{Department of Astronomy \& Astrophysics, Pennsylvania State University\\
525 Davey Laboratory, University Park, PA 16802, USA}
\altaffiltext{3}{Chandra Fellow}
\altaffiltext{4}{Institute of Space and Astronautical Science, Japan Aerospace
Exploration Agency\\ 3-1-1 Yoshinodai, Sagamihara, Kanagawa 229-8510}
\altaffiltext{5}{Department of Physics, Graduate School of Science, Kyoto University,
Kita-shirakawa Oiwake-cho, Sakyo, Kyoto 606-8502}
\altaffiltext{6}{Faculty of Humanities and Social Sciences, Iwate University, 3-18-34
Ueda, Morioka, Iwate 020-8550}
\email{miura@phys.chuo-u.ac.jp}
\KeyWords{stars: flare --- stars: magnetic fields --- X-rays: individual (HD\,161084)}
\maketitle

\begin{abstract}
 We report a serendipitous detection of an intense X-ray flare from the Tycho reference
 source HD\,161084 during a Suzaku observation of the Galactic Center region for
 $\sim$20~ks. The X-ray Imaging Spectrometer (XIS) recorded a flare from this A1-type
 dwarf or subgiant star with a flux of
 $\sim$1.4$\times$10$^{-12}$~erg~s$^{-1}$~cm$^{-2}$ (0.5--10~keV) and a decay time scale
 of $\sim$0.5~hr. The spectrum is hard with a prominent Fe\emissiontype{XXV} K$\alpha$
 emission line at 6.7~keV, which is explained by a $\sim$5~keV thin-thermal plasma
 model attenuated by a $\sim$1.4$\times$10$^{21}$~cm$^{-2}$ extinction. The low
 extinction, which is consistent with the optical reddening, indicates that the source
 is a foreground star toward the Galactic Center region. Based on the spectroscopic
 parallax distance of $\sim$530~pc, the peak X-ray luminosity amounts to
 $\sim$1$\times$10$^{32}$~erg~s$^{-1}$ (0.5--10~keV). This is much larger than the X-ray
 luminosity of ordinary late-type main-sequence stars, and the X-ray emission is
 unattributable to a hidden late-type companion that comprises a wide binary system with
 the A-star. We discuss possible natures of HD\,161084 and suggest that it is most
 likely an interacting binary with elevated magnetic activity in the companion such as
 the Algol-type system. The flux detected by Suzaku during the burst is $\sim$100 times
 larger than the quiescent level measured using the archived XMM-Newton and Chandra
 data. The large flux amplification makes this star a unique example among sources of
 this class.
\end{abstract}

\section{Introduction}
Main-sequence stars with intermediate (late B to early A) spectral types are considered
intrinsically X-ray inactive. These stars have no X-ray production mechanisms such as
shocks in unstable stellar winds causing the X-ray emission in early-type stars (earlier
than B2; \cite{lucy80,berghoefer97}), or the magnetic activity as a consequence of
surface convection layer and differential rotation responsible for the X-ray emission in
late-type stars. Indeed, most X-ray surveys of A-type stars in the field and in open
clusters show a consistent paucity of their X-ray detections
(e.g. \cite{schmitt85,schmitt90,micela90}).

X-ray emission from A-type stars, if detected, is usually attributed to its
magnetically-active late-type companion (e.g. \cite{huelamo00,briggs03}) that comprises
a binary system. The idea is consistent with the following observational facts: (1)
There is a lack of any correlation between the X-ray luminosity and various stellar
properties of A-type stars \citep{simon95,panzera99}. (2) The X-ray emission from A-type
stars is characterized by X-ray luminosities of 10$^{28}$--$10^{30}$~erg~s$^{-1}$ with
averaged soft emission, which is similar to those seen in late-type main-sequence stars.

In some high-resolution X-ray imaging studies, the position of the X-ray emission is
located precisely enough to claim that the emission is from the late-type companion and
not from the A-type star itself \citep{stelzer03a,stelzer03b}. In some unresolved
systems, too, the X-ray emission is confirmed to be from the late-type companion by the
eclipsing observation \citep{schmitt93} or by the Doppler shift measurements of emission
lines in the X-ray spectrum \citep{chung04}.

Nevertheless, these pieces observational evidence do not entirely rule out the
possibility of the intrinsic X-ray emission from A-type stars. A-type stars with a
debris disk (e.g., $\beta$ Pic) may radiate thermal X-rays fueled by accretion
\citep{hempel05} and those with strong magnetic activity (e.g, IQ Aur) may cause X-ray
emission caused by magnetically-confined winds \citep{babel97}.

The long-term X-ray behavior of A-type stars is an important property to characterize
their emission and to understand its origin. Monitoring observations on particular
A-type stars are necessary to reveal their long-term flux and spectral variation, but
the number of such studies is only a few (e.g. \cite{briggs03}). This is because
systematic X-ray studies of A-type stars have been conducted by utilizing all-sky survey
data \citep{simon95,panzera99} or by snapshot observations of open clusters including
A-type stars \citep{briggs03}.

\medskip

Here, we study the long-term X-ray behavior of an A-type star by exploiting the wealth of
data obtained in the Galactic Center region. Every X-ray satellites repeatedly observed
this region, which contains many A-type stars as foreground objects. We surveyed the
Suzaku images of the Galactic Center and serendipitously found that the A-type
star HD\,161084 showed an intense flare during one of the mapping observations. The star
is a Tycho reference source classified as A1\,V or A1\,IV \citep{wright03}. In the X-ray
band, the star has been soft and faint at $\sim$10$^{29.5}$~erg~s$^{-1}$ in the
0.5--10~keV luminosity in the archived X-ray data, but a sudden flaring made this source
as bright as $\sim$10$^{32}$~erg~s$^{-1}$, which is hard to achieve by a hidden ordinary
late-type companion. We report the result obtained by the X-ray Imaging Spectrometer (XIS)
on-board Suzaku and discuss the nature of this source.

\section{Observation}
We conducted a Suzaku observation centered at (R.\,A.,
decl.)\,$=$\,(\timeform{17h44m55.8s}, \timeform{-29D49'16''}) in the equinox J2000.0 as
a part of the Galactic Center mapping campaign during the second announcement of
opportunity observing cycle. The position is $\sim$50\arcmin\ apart from the Galactic
Center in the southwest direction. The observation was conducted on 2007 March 13--14
for a telescope time of $\sim$34.5~ks.

The Suzaku satellite \citep{mitsuda07} produces simultaneous data sets taken by two
instruments; one is the XIS \citep{koyama07} sensitive in the energy range below
$\sim$12~keV and the other is the Hard X-ray Detector \citep{kokubun07,takahashi07}
sensitive in the higher energy band. We use the XIS data in this paper.

XIS is equipped with four X-ray CCDs (XIS0--3) mounted at the focal planes of the four
independent X-Ray Telescopes \citep{serlemitsos07} aligned to observe a
$\sim$18\arcmin$\times$18\arcmin\ region. One of them (XIS1) is a back-side illuminated
(BI) CCD chip and the remaining three (XIS0, XIS2, and XIS3) are front-side illuminated
(FI) chips. The BI and FI chips are composed of 1024$\times$1024~pixels and are superior
to each other in the soft and hard band responses, respectively. One of the FI chips
(XIS2) turned dysfunctional in November 2006. We therefore use the data obtained by the
remaining three CCDs.

The capability of XIS is subject to degradation by charged particle radiation in the
orbit. XIS employs the spaced-row charge injection (SCI) technique to rejuvenate its
spectral resolution by filling the charge traps with artificially injected electrons
through CCD readouts. We used the SCI mode in this observation. The energy resolution as
of the observation date is $\sim$130~eV in the full width half maximum at 6~keV. The
radioactive sources of \atom{Fe}{}{55} illuminate two corners of each CCD for the
calibration of the absolute energy gain at an accuracy of $\sim$10~eV for the FI data
and $\sim$60~eV for the BI data in the SCI mode. The imaging performance is
characterized by a half power diameter of the point spread function of $\sim$2\arcmin,
which is independent of the off-axis angle. The absolute astrometry of the XIS frame is
uncertain up to $\sim$50\arcsec.

The observation was conducted with the normal clocking mode with a frame time of 8~s. We
reduced the data using HEAsoft\footnote{See
http://heasarc.gsfc.nasa.gov/docs/software/lheasoft/ for detail.} version 6.2.0. A
cleaned event list was obtained from the data (processing version 1.3\footnote{See
http://www.astro.isas.jaxa.jp/suzaku/process/ for details.}) by removing events taken
during the South Atlantic Anomaly passages, the geomagnetic cut-off rigidity of $<$8~GV,
the elevation angles from the Earth rim of $<$10$^{\circ}$ and from the sun-lit Earth
rim of $<$20$^{\circ}$, and the telemetry saturation. After the filtering, the net
integration time is $\sim$19.7~ks. The observation log is summarized in
table~\ref{tb:t1}.

\begin{table}
 \begin{center}
  \caption{Observation log.}\label{tb:t1}
  \begin{tabular}{llcr}
   \hline
   \hline 
   Telescope/ & Sequence & Observation & $t_{\rm{exp}}^{*}$ \\
   Instrument  & number   & date        & (ks)          \\
   \hline
   Einstein/IPC   & 2516         & 1979/09/21 & 1.8 \\
   ROSAT/PSPC     & rp400275     & 1992/09/28 & 15.1 \\
   ASCA/GIS       & 55038000     & 1997/03/20 & 30.7 \\
   Chandra/ACIS-S & 658          & 2000/08/30 & 9.2 \\
   Chandra/ACIS-I & 2278         & 2001/07/20 & 11.6 \\
   Chandra/ACIS-S & 2286         & 2001/07/21 & 11.6 \\
   Newton/MOS     & 0303210201   & 2005/10/02 & 23.4 \\
   Suzaku/XIS     & 501056010    & 2007/03/13 & 19.7 \\
   \hline
   \multicolumn{4}{@{}l@{}}{\hbox to 0pt{\parbox{85mm}{\footnotesize
   \par\noindent
   \footnotemark[$*$] Net exposure time.
  }\hss}}
  \end{tabular}
 \end{center}
\end{table}

\section{Results}
\subsection{Image and Source Identification}
\begin{figure*}[ht!]
 \begin{center}
  \FigureFile(170mm,85mm){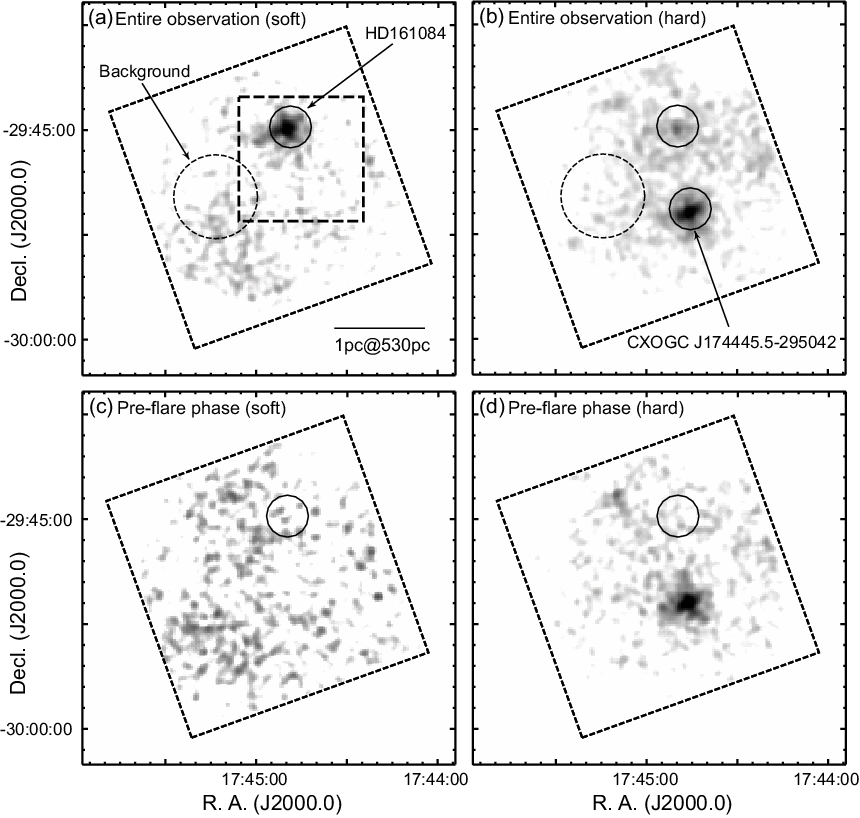}
 \end{center}
 \caption{Smoothed XIS images in the (a) 0.5--2.0~keV and (b) 2.0--10~keV band
 integrated over the entire observation and (c) 0.5--2.0~keV and (d) 2.0--10~keV band
 during the pre-flare phase (phase 1 in figure~\ref{fg:f3}). The signals of the three
 CCD chips are summed.  The calibration sources at the corners are masked. The two
 identified sources are labeled in (a). The extraction regions of the source and
 background signals for HD\,161084 are shown by solid and dotted circles,
 respectively. The close-up XMM-Newton view of the square dashed region in (a) is shown
 in figure~\ref{fg:f2}.}\label{fg:f1}
\end{figure*}

Figure~\ref{fg:f1} (a) and (b) show the XIS images integrated over the entire
observation in the (a) soft (0.5--2.0~keV) and (b) hard (2.0--10~keV) band. Both the FI
(XIS0 and 3) and BI (XIS1) data are summed. Two point-like sources were detected; one is
conspicuous in both bands at the top right and the other in the hard band near the center.

In order to register the XIS frame and to identify these two sources, we retrieved the
archived data of other X-ray telescopes. We found that Einstein, ROSAT, ASCA, Chandra,
and XMM-Newton covered this region during their observations of Galactic Center sources
(table~\ref{tb:t1}). We found that the Suzaku source near the field center was also
detected in the Chandra and at the edge of the XMM-Newton images with a similar spectral
hardness and brightness. The Chandra position of this source is registered using the
2MASS--Chandra counterpart matches in this region and is named CXOGC\,J174445.5--295042
\citep{muno06}. This source is point-like and isolated, thus serves as a good
astrometric calibrator. We shifted the astrometry of the XIS image to match with the
Chandra position.

\begin{figure}
 \begin{center}
  \FigureFile(85mm,85mm){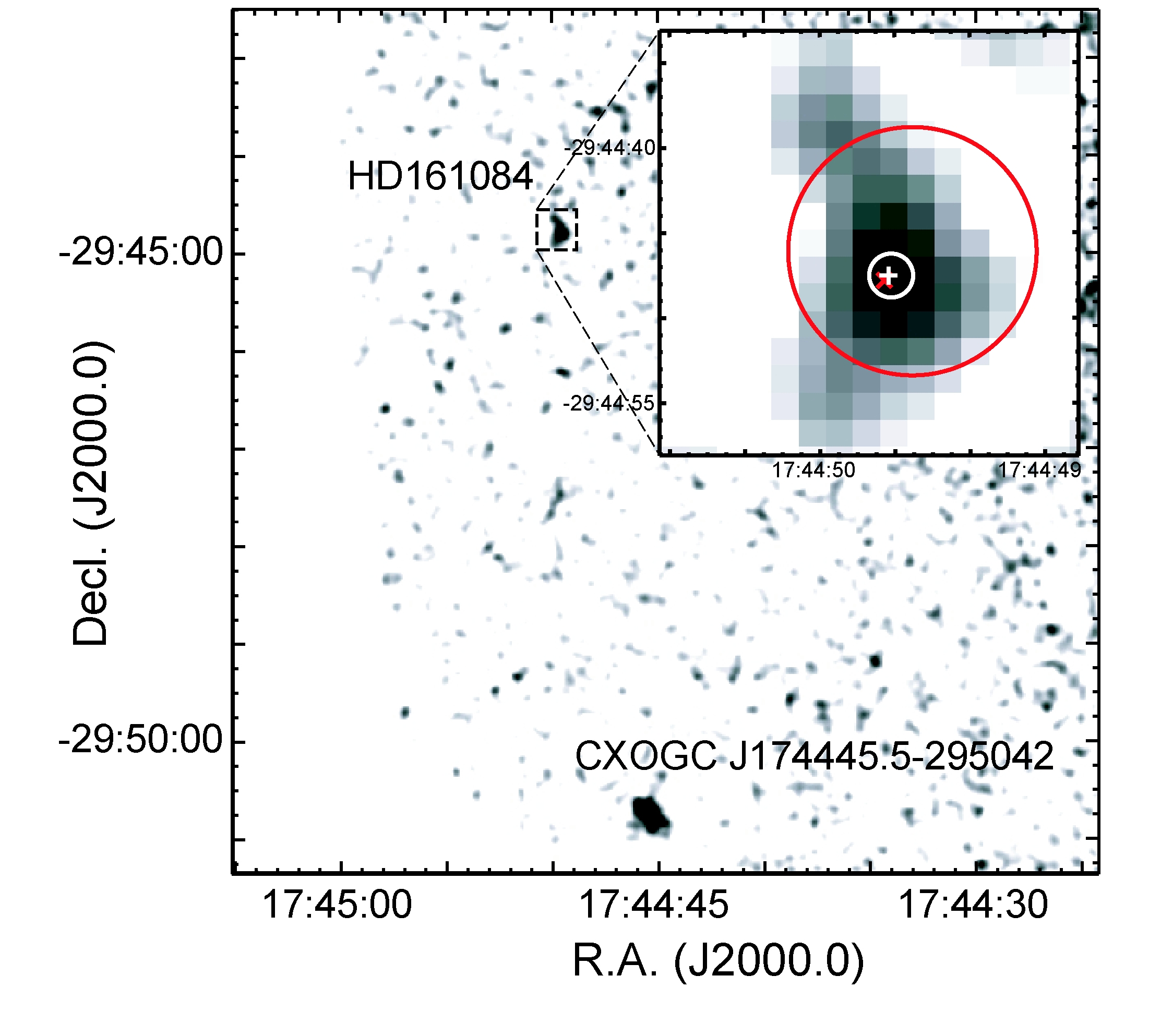}
 \end{center}
 \caption{XMM-Newton 0.5--10~keV band image of the dashed square region in
 figure~\ref{fg:f1}(a). Two MOS images are combined. The pn data were unavailable. The
 position of the Chandra \citep{muno06} and the Tycho sources are shown with a red cross
 and a white plus, respectively. The error circles (90\% confidence range) of the
 position are shown for the Suzaku and XMM-Newton counterparts of HD\,161084
 respectively by red and white circles.}\label{fg:f2}
\end{figure}

Using the registered XIS frame, we measured the position of the other Suzaku source to be
(R.\,A., decl.)\,$=$\,(\timeform{17h44m49.57s}, \timeform{-29D44'46.1''}). The
positional uncertainty is $\sim$7\farcs3 (90\% confidence range), which includes both
the uncertainty of the position determination by fitting the intensity profile of the
source and the uncertainty of the plate scale of the XIS image. Within the Suzaku
error circle, we found a Chandra source cataloged as CXOGC\,J174449.7--294447 and
identified as the X-ray counterpart of HD\,161084 \citep{muno06}. We confirmed that
there is no other source in the vicinity in the Chandra image as well as
in near-infrared
images obtained by SIRIUS (Simultaneous three-color InfraRed Imager for Unbiased
Surveys; \cite{nagayama03}) at the Infrared Survey Facility (private communication with
S. Nishiyama and T. Nagata). We thus conclude that the Suzaku source is the X-ray
counterpart of HD\,161084 and is identical to CXOGC\,J174449.7--294447.

We also checked the XMM-Newton image obtained by the European Photon Imaging Camera
(EPIC; \cite{strueder01,turner01}). The profile of the astrometric calibrator is
distorted at the field edge and could not be used for the field alignment. No other
sources in the image were suitable for this purpose, so we did not perform any
bore-sight correction. Nonetheless, the astrometry given in the processed
data\footnote{See http://xmm.esac.esa.int/docs/documents/CAL-TN-0018-2-5.pdf for
details.} has uncertainty of $\sim$3\farcs6 (90\% confidence range), which is sufficient
to compare to the other images. Figure~\ref{fg:f2} shows the close-up view of the
XMM-Newton image. A source was confirmed at the position of HD\,161084, which we
recognize as the XMM-Newton counterpart of HD\,161084.

Although HD\,161084 was detected in the Suzaku, Chandra, and XMM-Newton images, their
X-ray flux is quite different; the Chandra and XMM-Newton flux is fainter than the
Suzaku flux by about two orders. Actually, the source was faint and undetected in the
first 10~ks (phase 1 in figure~\ref{fg:f3}; see \S3.2) of the Suzaku observation either,
which is evident in the time-sliced XIS images (figure~\ref{fg:f1}) of this duration in
the (c) soft and (d) hard band. This indicates that the source experienced a flux
amplification during the XIS observation.

\subsection{Light Curve}
\begin{figure}
 \begin{center}
  \FigureFile(85mm,85mm){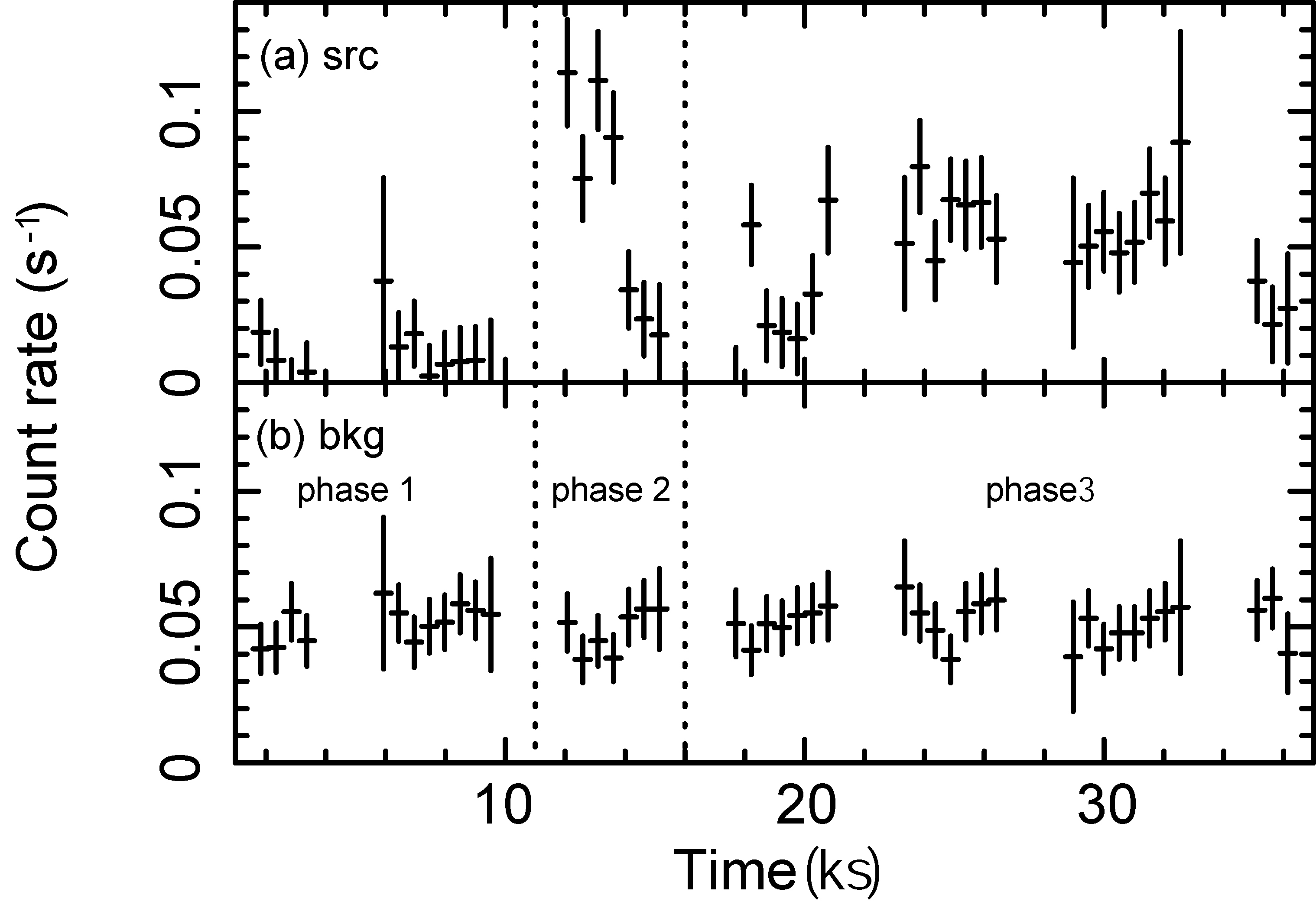}
 \end{center}
 \caption{(a) Background-subtracted light curve of HD\,161084 and (b) the background
 light curve normalized to the extraction area in the 0.5--10~keV band. The three XIS
< data were summed. The abscissa shows the time from the observation start. Three
 time-slices (phase 1--3) are defined based on the variation of the source count
 rate.}\label{fg:f3}
\end{figure}

We constructed the X-ray light curve of HD\,161084 in the 0.5--10~keV band using both
the FI and BI data. The source photons were accumulated from a circular region with a
radius of 1\farcm5, while the background photons were from a circular region with a
3\farcm0 radius (solid and dashed circles in figure~\ref{fg:f1}a, respectively). The
centers of the two circles are at an equal distance from CXOGC\,J174445.5--295042 in
order to cancel a possible contamination from the source. Figure~\ref{fg:f3} shows the (a)
background-subtracted light curve and (b) background light curve normalized to the
extraction area.

In the source light curve, we see a variation of a rapid rise and a slow decay typical
of stellar flares starting $\sim$12~ks after the observation started. The variation is
hardly attributable to the background, as the background light curve is relatively
stable. The duration of the flare is characterized by an $e$-folding time of
$\sim$1.7$^{+0.9}_{-0.5}$~ks. After the flare ceases at $\sim$16~ks, the count rate did
not settle back to the pre-flare quiescent phase, but it subsequently showed another
flux amplification that lasted until the end of the observation. We defined three time
slices based on this development: phase 1 for the pre-flare phase, phase 2 for the flare
phase, and phase 3 for the post-flare phase.

\subsection{Spectrum}
\begin{table*}
 \begin{center}
  \caption{Best-fits spectral parameters.\footnotemark[$*$]}\label{tb:t2}
  \begin{tabular}{lcccccc}
   \hline
   \hline 
   Telescope/Instrument & $N_{\rm H}$ & $k_{\rm{B}}T$ & $EM$\footnotemark[$\dagger$] &
   $F_{\rm{X}}$\footnotemark[$\ddagger$] & $L_{\rm{X}}$\footnotemark[$\dagger\ddagger$]&
   $\chi^{2}$/d.o.f \\
   & ($10^{22}$~cm$^{-2}$) & (keV) & (10$^{53}$~cm$^{-3}$) 
   & (erg~s$^{-1}$~cm$^{-2}$) & (erg~s$^{-1}$) & \\
   \hline
   Suzaku/XIS (phase2) & 0.22$^{+0.35}_{ -0.20}$ & 5.2$^{+5.9}_{-2.0}$ 
	   & 25.3 $^{+9.3}_{-6.9}$ & 13.5$^{+3.6}_{-3.6} \times 10^{-13}$~& 4.6$\times$10$^{31}$ & 18.9/18  \\
   Suzaku/XIS (phase3) & 0.11$^{+0.14}_{ -0.10}$ & 4.0$^{+2.8}_{-1.4}$
	   & 15.6 $^{+3.3}_{-2.9}$ & \phantom{0}7.7$^{+1.7}_{-1.6} \times 10^{-13}$~& 2.6$\times$10$^{31}$ & 15.3/20  \\
   Suzaku/XIS (phase2+3) & 0.14$^{+0.11}_{ -0.09}$ & 4.7$^{+2.5}_{-1.3}$
	   & 18.0 $^{+3.0}_{-2.7}$ & \phantom{0}9.3$^{+1.5}_{-1.2} \times 10^{-13}$~& 3.2$\times$10$^{31}$ & 22.3/20  \\
   \hline
   XMM-Newton/MOS       & 0.14 (fixed)            & 0.66$^{+0.47}_{-0.35}$
	   & 0.19$^{+0.09}_{-0.09}$ & \phantom{0}1.5$^{+0.5}_{-0.7} \times 10^{-14}$~& 4.9$\times$10$^{29}$ & 8.44/10 \\
   \hline
   \multicolumn{4}{@{}l@{}}{\hbox to 0pt{\parbox{190mm}{\footnotesize
   \par\noindent
   \footnotemark[$*$] The uncertainties indicate the 90\% confidence range.
   \par\noindent  
   \footnotemark[$\dagger$] A distance of 530~pc is assumed (see \S~\ref{sect:4-1}).
   \par\noindent
   \footnotemark[$\ddagger$] Values in the 0.5--10~keV energy range.
    }\hss}}
  \end{tabular}
\end{center}
\end{table*}

\begin{figure}
 \begin{center}
  \FigureFile(85mm,85mm){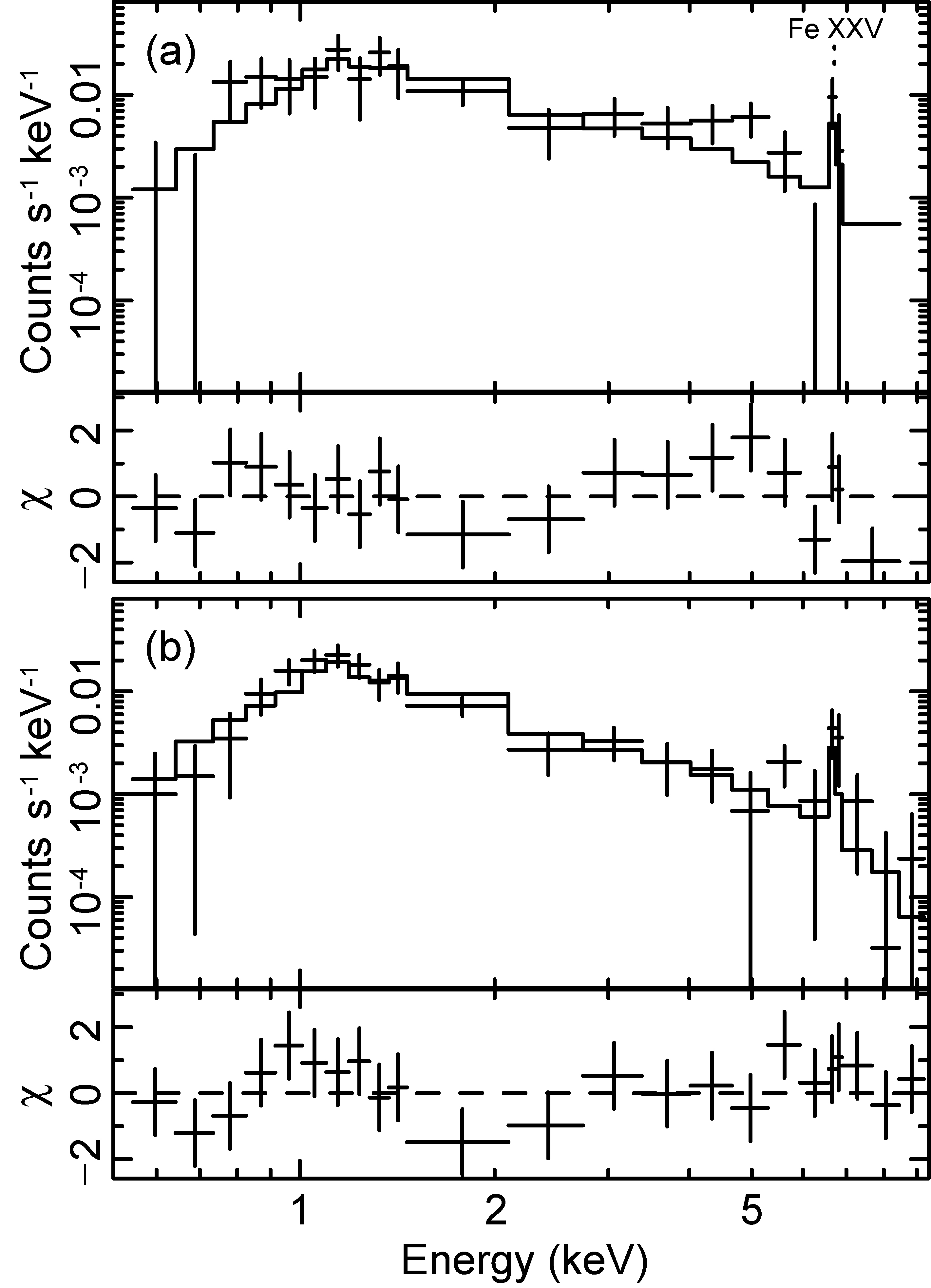}
 \end{center}
 \caption{Background-subtracted XIS spectra in the (a) phase 2 and (b) phase 3
 (figure~\ref{fg:f3}). Data obtained by the two FI chips were merged. The top panel shows the data with crosses and the best-fit model
 with solid lines. The bottom panel shows the residuals to the fit.}\label{fg:f4}
\end{figure}

We next constructed the X-ray spectrum of HD\,161084. We used a provisional
method\footnote{See http://www.astro.isas.jaxa.jp/suzaku/analysis/xis/sci/ for detail.} 
for the spectral analysis of the SCI data. This method is only applicable to the FI
data, so we did not use the BI data in the spectral analysis. Figure~\ref{fg:f4} shows
the background-subtracted spectra in the 0.5--10~keV band during the (a) phase 2 and (b)
phase 3. The source and the background spectra were accumulated from the same regions
for the light curves.

Both spectra are characterized by hard emission with a conspicuous emission line feature
at 6.7~keV, which originates from the K$\alpha$ line of Fe\emissiontype{XXV}. It is evident
that the spectra are thermal with a plasma temperature of a few keV. We fitted the data
with a thin-thermal plasma (\texttt{mekal}) model
\citep{mewe85,mewe86,kaastra92,liedahl95} with an interstellar extinction
\citep{morrison83}. The abundance of elements was fixed to the solar values
\citep{anders89}. Here, we generated ancillary responses using \texttt{xissimarfgen}
(version 2006-11-26; \cite{ishisaki07}) and the redistribution matrices using the
provisional method.

The best-fit values of the plasma temperature ($k_{\rm{B}}T$), X-ray volume emission
measure ($EM$), flux ($F_{\rm{X}}$), and luminosity ($L_{\rm{X}}$) and the amount of
extinction ($N_{\rm{H}}$) are summarized in table~\ref{tb:t2}. A
statistically-acceptable fit was obtained for both spectra. Upon the confirmation that
the spectral shape did not change between the phase 2 and 3, we merged the two spectra
and derived the best-fit parameters for the integrated spectrum (table~\ref{tb:t2}). In
order to examine how the best-fit parameters are affected by our choice of the
background region, we repeated the same procedure for several other background regions
and found that the best-fit values were consistent with each other.

\subsection{Long-term Behavior}
\begin{figure}
 \begin{center}
  \FigureFile(85mm,85mm){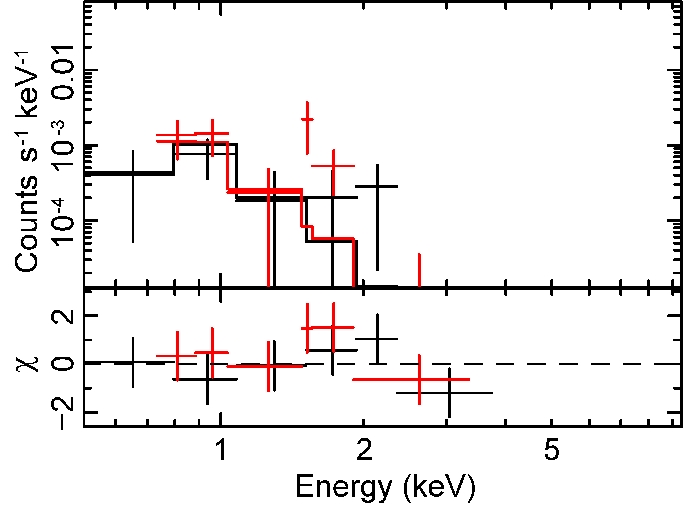}
 \end{center}
 \caption{Background-subtracted EPIC spectra (black for MOS1 and red for MOS2). Other
 symbols follow figure~\ref{fg:f4}.}\label{fg:f5}
\end{figure}

Using the archived data, we studied the long-term X-ray variation of this source. Out of
the seven archived data sets with six X-ray telescopes for the last 30 years
(table~\ref{tb:t1}), only one observation by XMM-Newton yielded sufficient counts for
spectral analysis. We constructed an X-ray spectrum using the EPIC MOS data
(figure~\ref{fg:f5}). We fixed the extinction value to that obtained by the XIS
($N_{\rm{H}} = 1.4 \times 10^{21}$~cm$^{-2}$) and fitted the data with the same
model. The best-fit values (table~\ref{tb:t2}) are distinctively different from the
Suzaku results; the plasma temperature is lower by an order and the luminosity is
smaller by two orders than those in the phases 2 and 3 of the Suzaku observation.

\begin{figure}
 \begin{center}
  \FigureFile(85mm,85mm){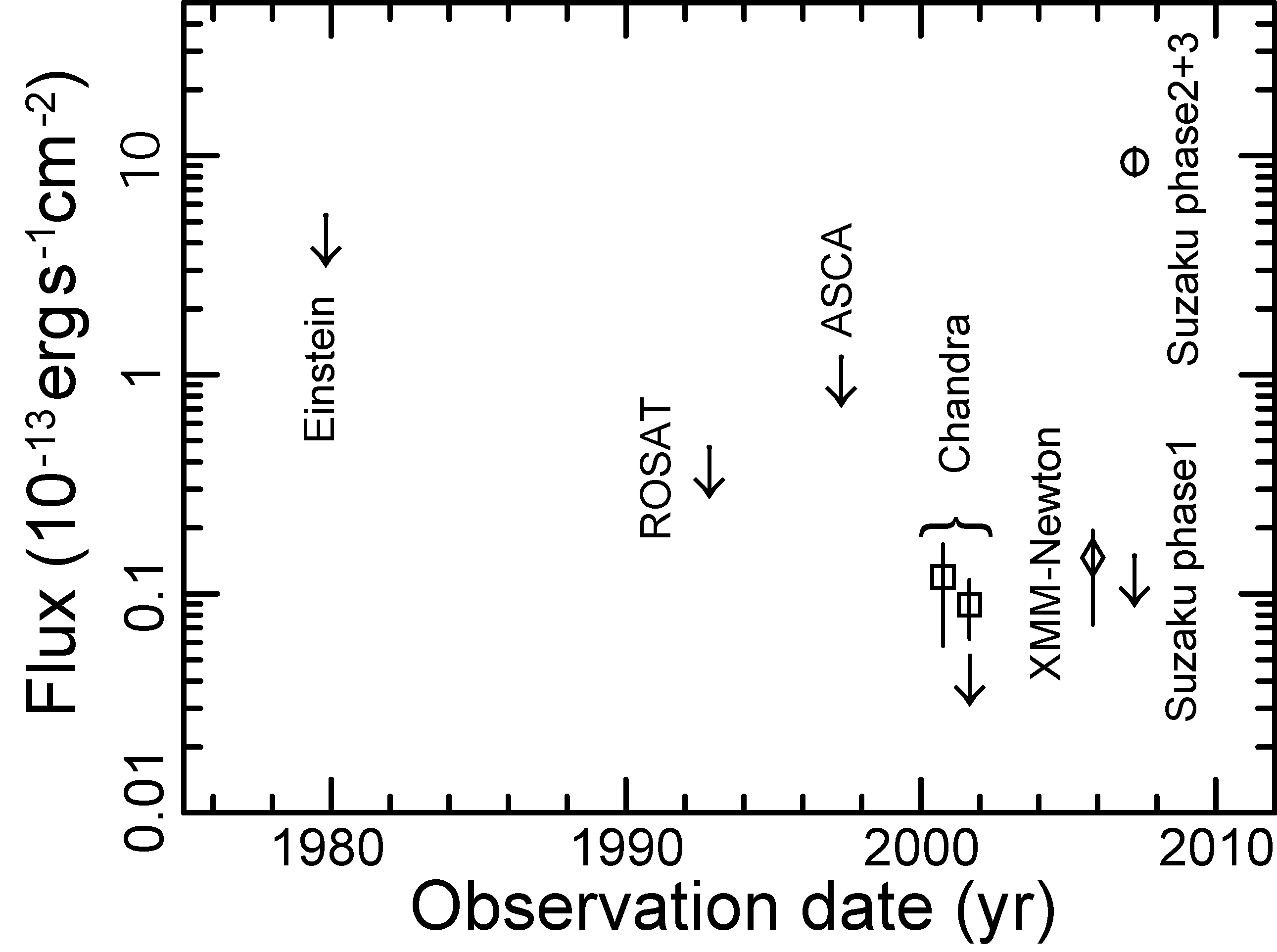}
 \end{center}
 \caption{Long-term flux variation of HD\,161084 in the 0.5--10~keV band. The X-ray
 emission was detected in two of the three Chandra (squares), one XMM-Newton (diamond),
 and the phases 2 and 3 Suzaku (circle) observations. The bars indicate the uncertainty
 of the flux measurements. The upper limits (downward-pointing arrows) were obtained for
 the remaining observations and phase.}\label{fg:f6}
\end{figure}

Other data sets were either too poor for spectral fits (two Chandra observations;
ObsID\,$=$\,658 and 2278) or with no significant X-ray detections at all. We therefore
assumed the spectral shape of the best-fit XMM-Newton model and derived the flux or its
3~$\sigma$ upper limit in the 0.5--10~keV band. The flux upper limit in the Suzaku phase
1 was also derived in the same manner. The long-term flux variation is shown in
figure~\ref{fg:f6}. During the Suzaku observation, the flux was amplified by more than
two orders from the quiescent level determined by the Newton and Chandra
observations. No such burst event was detected in other observations.

\section{Discussion}
\subsection{Extinction and Distance to HD\,161084}\label{sect:4-1}
HD\,161084 has been a largely ignored source and nothing beyond the catalog information
is known. We start with estimating the distance to HD\,161084. From the SIMBAD database,
the source has a \textit{B}-band and \textit{V}-band magnitudes of 10.29~mag and
10.11~mag, respectively. It is classified as a star of A1\,VI or A1\,V.

Assuming the source is an A1\,V star, the intrinsic (\textit{B}--\textit{V}) color is
$\sim$0.015~mag \citep{drilling00}. The source is estimated to have a reddening of
(\textit{B}--\textit{V}) $\sim$~0.165~mag to be compatible with the observed color,
which can be converted to an extinction of $A_{V} \sim$~0.51~mag. This is consistent
with the independent measurement of the X-ray absorption of
$\sim$1.4$\times$10$^{21}$~cm$^{-2}$ assuming the conversion factor of
$N_{\rm{H}}$/$A_{V}=$1.79$\times$10$^{21}$~cm$^{-2}$~mag$^{-1}$ \citep{predehl95}. The
agreement gives another support for the identification of the Suzaku source with
HD\,161084. From the dereddended \textit{V}-band magnitude of HD\,161084 and the
intrinsic \textit{V}-band magnitude of A1\,V stars, we estimate a distance of
$\sim$530~pc.

We repeated the same procedure assuming that HD\,161084 is an A0\,V or A2\,V star to see
how the uncertainty of the spectral-type measurement affects our distance estimate. The
intrinsic (\textit{B}--\textit{V}) color differs by $\sim$0.035~mag and the absolute
\textit{V}-band magnitude by $\sim$0.33~mag among A0\,V, A1\,V, and A1\,V types
\citep{drilling00}. The derived distance is $\sim$586~pc for A0\,V and $\sim$480~pc for
A2\,V. This converts to the uncertainty of the X-ray luminosity by $\sim$20\%.

\subsection{Flaring and Quiescent X-rays}\label{sect:4-2}
We detected intense X-ray emission from HD\,161084 with a flare-like light curve and
hard emission accountable by a plasma temperature of $\sim$5~keV. The peak luminosity at
the maximum count rate bin in figure~\ref{fg:f3} is
$\sim$1$\times$10$^{32}$~erg~s$^{-1}$. The $e$-folding time of the star is similar to
those of other flare stars such as the Sun and RS CVn-type binaries. Based on the plot
of X-ray volume emission measure versus the plasma temperature \citep{shibata99}, we
estimate the magnetic field strength to be $\sim$50~gauss and the flare loop length to
be $\sim$0.05~AU. The flare was detected once in a total duration of $\sim$9 days
(table~\ref{tb:t1}), which indicates that the flare frequency is once in 2--113 days
(90\% confidence; \cite{kraft91}).

The XMM-Newton spectrum was obtained during a quiescent state of this source. We could
not obtain any meaningful X-ray spectra from other observations and from the first 10~ks
(phase 1) of the Suzaku observation. We speculate that HD\,161084 was at a quiescent
state in these observations with a similar X-ray brightness and hardness recorded by
XMM-Newton.

\subsection{Nature of HD\,161084}\label{sect:4-3}
In the quiescent phase, the X-ray emission from HD\,161084 appears to be attributable to
an uncovered normal low-mass main-sequence companion or the intrinsic emission from an
A-type star. However, the intense flare with an amplified X-ray flux and a hard spectrum
makes both interpretations unlikely. The flare luminosity is beyond the reach of normal
late-type stars. The X-ray emission from the debris disk and magnetic A-type stars is
weak ($\sim$10$^{28}$~erg~s$^{-1}$), soft, and constant \citep{hempel05,cash82,golub83},
which contradicts the behavior of HD\,161084 during the flare.

We speculate that the X-ray emission from HD\,161084 is from its hidden late-type
companion, and the secondary star is exceptionally bright among stars of this type. The
level of X-ray activity in late-type stars is known to be an increasing function of
rotational velocity (e.g. \cite{pallavicini81}). This is a natural consequence that the
X-ray--emitting plasma is fueled by the energy release of magnetic fields generated by
the turbulence in the surface convection layer and the differential rotation. Two major
classes of late-type stars are known to rotate faster and emit exceptionally brighter
X-rays than ordinary late-type stars.

One is pre--main-sequence sources, whose faster rotation stems from the accumulation of
angular momentum via accretion and contraction during their formation. These sources can
be bright up to $\sim$10$^{32}$~erg~s$^{-1}$ with hard X-ray emission (e.g,
\cite{imanishi01}). A bright and hard X-ray flare of $\sim$10$^{31.6}$~erg~s$^{-1}$
(0.5--8.0~keV) was observed from an intermediate-type star HD\,38563S (B3--B5;
\cite{yanagida04}). The X-ray emission from this source can be from a hidden late-type
pre--main-sequence companion around the B star, which is quite conceivable considering
that HD\,38563S is in a young star-forming region NGC\,2068. We consider that
HD\,161084, in contrast, is not located in a star-forming region and thus is unlikely to
have a pre--main-sequence companion. We examined the near-infrared image by SIRIUS, but
no clustering of red sources was present, which would indicate star-forming
regions. Moreover, the extinction toward HD\,161084 is consistent with the integration
of interstellar hydrogen density of 1~cm$^{-3}$ along the line of sight. Most sources in
star-forming regions suffer an additional extinction by natal molecular cloud and
circumstellar matter. The lack of such extra extinction further supports that HD\,161084
is a field star.

The other class of fast rotators is interacting binaries. The tidal force and mass
transfer between the two components of a close binary synchronizes the stellar rotation
with the orbital motion, giving rise to a faster rotation and elevated X-ray activity in
the late-type constituent. A handful of classes of such interacting binaries are known
\citep{hall89,richards93}. Among them, the classes that may contain an intermediate-type
main-sequence star are the Algol-type (semi-detached) and the W UMa-type (contact)
binaries.

The Algol-type binary is a well-established class of intense X-ray emitters with an
intermediate-type dwarf and late-type giants or sub-giants. The X-ray emission from
Algol-type sources are attributed to fast-rotating late-type star
\citep{white83,schmitt99,chung04}. They emit hard and luminous X-rays with occasional
flares \citep{singh95,singh96,ottmann96}, which is comparable to those observed from
HD\,161084 in the flare state. We suggest that HD\,161084 is an interacting binary of
the Algol-type based on the observed X-ray properties.

W UMa-type stars also cause X-ray flares with the luminosity up to
$\sim$10$^{31}$~erg~s$^{-1}$ \citep{mcgale96,choi98}. In almost all these systems, the
spectral type of the two stars are nearly equal \citep{maceroni96}. The secondary star
of HD\,161084 is thus likely to be another A1-type star. The origin of the X-ray
emission boils down to the intrinsic X-ray emission from A-type stars, which we
concluded unlikely to account for the emission from HD\,161084. We thus speculate that
HD\,161084 is not a W UMa-type binary.

What makes this source unique among these classes is the degree of amplification of the
flare flux from the quiescent flux. Both the Algol and W UMa binaries show occasional
flares, but their flux increase from the quiescent state is commonly a few times
\citep{choi98,gondoin04,gondoin04_2,mcgale96,nordon07,white86}. It may be the case that
the X-ray emitting late-type companion in the HD\,161084 system was eclipsed during all
the previous X-ray observations except for the phases 2 and 3 of the Suzaku
observation. The interacting binary nature of HD\,161084 is conclusively identified by
established methods, such as photometric eclipse and spectroscopic monitoring. We
encourage follow-up observation of this source.

\section{Summary}
We serendipitously detected an intense X-ray flare from the A1-type star HD\,161084
during one of the Suzaku mapping observations of the Galactic Center region. The flux of
$\sim 1.4 \times 10^{-12}$~erg~s$^{-1}$~cm$^{-2}$ (0.5--10 keV) was recorded. The
spectrum shows a prominent Fe\emissiontype{XXV} K$\alpha$ emission line at 6.7~keV,
indicating a high temperature plasma. The flare decay time scale is $\sim$0.5~hr. We
estimated the distance of HD\,161084 as $\sim$530~pc based on the spectroscopic
parallax. The peak luminosity of the flare amounts to $\sim$10$^{32}$~erg~s$^{-1}$
(0.5--10 keV).

The flare emission is too luminous and too hard to explain by an intrinsic emission from
A-type stars or a hidden late-type main-sequence companion in a wide binary. We suggest
that the nature of this source is most likely an interacting binary such as the
Algol-type system, in which the magnetic activity of the late-type companion is enhanced
by fast rotation due to locking with the orbital motion.

The position of this source at the Galactic Center allowed us to study its long-term
behavior by exploiting the wealth of the archived data in this region. The quiescent
level was characterized by an X-ray luminosity of $\sim 10^{29.5}$~erg~s$^{-1}$ and a
plasma temperature of $\sim$0.7~keV, which are distinctively different from those during
the flare.

HD\,161084 showed two orders of amplification in luminosity from the quiescent to the
flare phase. No Algol-type sources are found with such a large amplification before,
which makes HD\,161084 a unique sample of these possible classes.

\bigskip

The authors thank Dai Takei and Masashi Hanya for their help in the archived X-ray data
reduction, Shogo Nishiyama and Tetsuya Nagata for providing the SIRIUS data, and Shinya
Narusawa for expertise comments on interacting binary systems. This research has made
use of the SIMBAD database, operated at CDS, Strasbourg, France.  Y.\,T., Y.\,M., and
S.\,Y. acknowledge support from the Grants-in-Aid for Scientific Research (numbers
17740109, 17740126 and 18540228, respectively) provided by the Ministry of Education,
Culture, Sports, Science, and Technology of Japan.

\end{document}